\documentclass[10pt,conference]{IEEEtran}
\usepackage{psfrag,color}
\usepackage{subfigure}
\usepackage{epsf}
\usepackage[final]{graphicx}
\usepackage[final]{epsfig}
\usepackage{amssymb}
\usepackage{amsmath}
\usepackage{amsopn}
\usepackage{amsthm}
\usepackage{graphicx}
\usepackage{cite}
\usepackage{times}
\usepackage[T1]{fontenc}

\usepackage[ps2pdf,
bookmarks,
colorlinks,
linkcolor={blue},
citecolor={blue},
urlcolor={blue},
pdfpagelabels,
bookmarksnumbered=true,
pagebackref,
]{hyperref}

\newcommand{\rem}[1]{}
\newcommand{\sq}{n'}					% bit time index
\newcommand{\sk}{k}					% k-th layer
\newcommand{\sn}{n}					% time index
\newcommand{\sj}{l}					% bit label index for channel input

\newcommand{\sri}{y}
\newcommand{\sr}{\mathbf{\sri}}

\newcommand{\sxi}{x}
\newcommand{\sx}{\mathbf{\sxi}}

\newcommand{\swi}{w}
\newcommand{\sw}{\mathbf{\swi}}
\newcommand{\ssigw}{\sigma^2}

\newcommand{\sHi}{H}
\newcommand{\sH}{\mathbf{\sHi}}

\newcommand{\sMt}{M_\text{T}}		% # of transmit antennas
\newcommand{\sMr}{M_\text{R}}		% # of receive antennas

\newcommand{\sbit}{b}					% information bit
\newcommand{\scbit}{c}					% coded bit
\newcommand{\scj}{\scbit_{\sj}}	% coded bit

\newcommand{\sA}{\mathcal{A}}
\newcommand{\sM}{|\sA|}					% alphabet length
\newcommand{\sm}{m}					% alphabet bit length

\newcommand{\sLLR}{\Lambda}			% LLR
\newcommand{\sLLRj}{\sLLR_{\sj}}	% LLR
\newcommand{\slambda}{{R_0}}
\newcommand{\shAz}{\mathcal{X}^0_{ \sj }}
\newcommand{\shAo}{\mathcal{X}^1_{ \sj }}	
\newcommand{\srate}{R}

\newcommand{\define}{\triangleq} %{\stackrel{\scriptscriptstyle \triangle}{=}}

\newcommand{\be}{\begin{equation}}
\newcommand{\ee}{\end{equation}}
\newcommand{\bee}{\begin{equation*}}
\newcommand{\eee}{\end{equation*}}

\newcommand{\zbf}{\mathbf{z}}

\newcommand{\Abf}{\mathbf{A}}

\newcommand{\Ibf}{\mathbf{I}}

\newcommand{\Ubf}{\mathbf{U}}

\newcommand{\Sigmabf}{\mathbf{\Sigma}}

\newcommand{\Ic}{\mathcal{I}}

\newcommand{\CNc}{\mathcal{CN}}
\newcommand{\Qc}{\mathcal{Q}}

\graphicspath{{pics/}{plots/}}

\title{Quantization for Soft-Output Demodulators in Bit-Interleaved Coded Modulation Systems}

\author{\IEEEauthorblockN{~\\[-5mm]\it Clemens Novak, Peter Fertl, and Gerald Matz}
\IEEEauthorblockA{~\\[-2mm]Institut f\"ur Nachrichtentechnik und Hochfrequenztechnik, 
Vienna University of Technology\\
Email: \{cnovak, pfertl, gmatz\}@nt.tuwien.ac.at }}

\begin{document}

% make the title area
\maketitle

\begin{abstract}
We study quantization of log-likelihood ratios (LLR) in bit-interleaved coded modulation (BICM) systems in terms of an equivalent discrete channel. 
% We propose a simply implementable quantizer design that targets a uniform distribution of the quantizer output.
We propose to design the quantizer such that the quantizer outputs become equiprobable.
% which allows for simple implementation.
% This approach allows for an easily implementable quantizer. %and give guidelines on the choice of the quantization values.
%
%
%We provide a quantizer design based on a uniform distribution of quantization values (quantization intervals and values) which is easily implementable and investigate semi-analytically
We investigate semi-analytically and numerically the ergodic and outage capacity over single- and multiple-antenna channels for different quantizers.
Finally, we show bit error rate simulations for BICM systems with LLR quantization using a rate 1/2 low-density parity-check code.
\end{abstract}

\section{Introduction}

% \subsection{Background}

{\em Bit-interleaved coded modulation (BICM)} is an attractive scheme for wireless communications 
where a block of information bits is mapped to transmit symbols via a
channel encoder and a symbol mapper separated by a code bit interleaver \cite{Fabregas:2008}. %\cite{caire98}. 
At the receiver side, a demodulator (demapper, detector) calculates 
% soft information , usually in the form of 
{\em log-likelihood ratios (LLR)} for the code bits,
which are de-interleaved and passed to the channel decoder.
%
%The system capacity of the equivalent BICM (modulation) channel between a code bit 
% (at the input of the mapper)
%and its corresponding LLR 
% (at the demodulator output) 
%was studied in \cite{caire98}. This work was extended to sub-optimum demodulators % for BICM 
%in a {\em multiple-input multiple-output (MIMO)} context
%in \cite{Fertl:2008aa}. %, leading to capacity results of BICM with different detectors.
% The performance of BICM systems can be further improved by using soft detection, where the demodulator not only provides bit decisions, but also provides a reliability measure, usually a log-likelihood ratio (LLR), on these decisions.
% By appropriately using this reliability information in a channel decoder, significant performance gains can be achieved.
%

Theoretically, one real-valued LLR per code bit needs to be computed and stored by the receiver.
% and processed by the decoder. 
Clearly, practical digital implementations can only use finite word-length 
approximations of real numbers,
% per LLR,  LLR approximations represented  finite number of bits, 
% thus requiring suitable 
which motivates the study of LLR quantization. 
We note that LLR quantization is also relevant % 
% our approach is also applicable to relaying 
for wireless (relay) networks that perform distributed turbo and network coding by 
exchanging soft information between different nodes. 
Optimal LLR quantization maximizing information rate for the special case of BPSK modulation over an AWGN channel was considered 
in \cite{Rave:2009}. 
% a method to optimally quantize these LLRs with regard to information rate was presented, but only for AWGN channels. However, in practice an 
However, an extension of this % optimization 
approach to other channels and modulations appears infeasible.
Thus, we consider a different quantizer design in this paper which allows for simple implementation 
% leading to an easily implementable scheme 
while only slightly degrading information rate.
%
% In this paper we study LLR quantization in BICM systems from an information theoretic perspective and we assess the capacity when using quantized LLRs instead of the optimum LLRs.
% use these concepts and consider the capacity of the equivalent channel between a bit at the input and the corresponding quantized LLR at the output. 
% \subsection{Contributions}
% We consider BICM systems with soft-out demodulator and consider the effect of LLR quantization. 
% More specifically, 
% 
More specifically, our contributions are as follows:

\begin{itemize}
	\item We propose to quantize the LLRs such that the quantizer outputs become equiprobable and provide appropriate reliability information to the channel decoder. %provide design guidelines leading to an easily implementable quantizer.
	\item We investigate the impact of the proposed LLR quantization on ergodic and outage rate,
		 using a semi-analytical approach for single-input single-output (SISO) systems with BPSK and Monte-Carlo simulations otherwise. 
%		 We also compare the rates of our design with a rate-optimal quantizer (cf.\ \cite{Rave:2009}).
% 	\item For the case of a single-input single-output (SISO) fading channel with BPSK modulation, we present analytic results for the equivalent modulation channel and its system capacity; for MIMO fading channels and higher order modulation we provide numerical results.
	\item We develop a method for designing and implementing the proposed quantizer during data
	      transmission. % , i.e., without the need to pre-compute any quantizer parameters.
% 	``on the fly'' i.e.\ without the need for lookup tables to store quantization intervals and quantizer outputs.
	\item We provide bit error rate (BER) simulations for BICM systems with LLR quantization
	      using {\em low-density parity-check (LDPC)} codes. %, which compare different quantization schemes.

\end{itemize}

% relaying, with the relay quantizing the received symbols before transmitting them to the receiver.

% We investigate the impact of the number quantization levels and the design of the quantizer on the system capacity and BER performance of the BICM system.

The paper is organized as follows. Section \ref{sec.model} presents the system model and
Section \ref{sec.quant} discusses the proposed %rate-optimal 
LLR quantization based on an equivalent discrete channel.
In Sections \ref{sec.siso} and \ref{sec.mimo}, we study the system capacity of 
SISO- and MIMO-BICM systems, respectively. 
The estimation of the quantizer parameters is addressed in Section \ref{sec:estimator} 
and BER results are provided in Section \ref{ssec:ber}.
% Conclusions are provided in Section \ref{sec.conclusion}.

\section{System Model}
\label{sec.model}

\begin{figure*}
\centering
\small

	\psfrag{a}{$\sbit[\sq]$}
	\psfrag{b}{encoder}
	\psfrag{c}[c]{$\Pi$}
	\psfrag{d}{\hspace*{.8mm}$c_l\hspace*{-.15mm}[n\hspace*{-.1mm}]$}
	\psfrag{e}{DEMUX}
	\psfrag{f}{map.}

	\psfrag{g}{$\sx[\sn]$}

	\psfrag{h1}{MIMO}
	\psfrag{h2}{channel}

	\psfrag{j}{$\sr[\sn]$}	
	\psfrag{k}{demodulator}
	\psfrag{n}{$\sLLRj[n]$}
	\psfrag{m}{MUX}
	\psfrag{q}{$\Pi^{-1}$}
	\psfrag{o}{decoder}
	\psfrag{p}{$\hat\sbit[\sq]$}

	\psfrag{r}{$\Qc(\cdot)$}
	\psfrag{s}{$\,d_l[n]$}

	\psfrag{z}{equivalent discrete channel}

	\psfrag{b1}[c]{$p_l[n]$}
	\psfrag{b2}[c]{$\bar{p}_l[n]$}

	\resizebox{17cm}{!}{\includegraphics{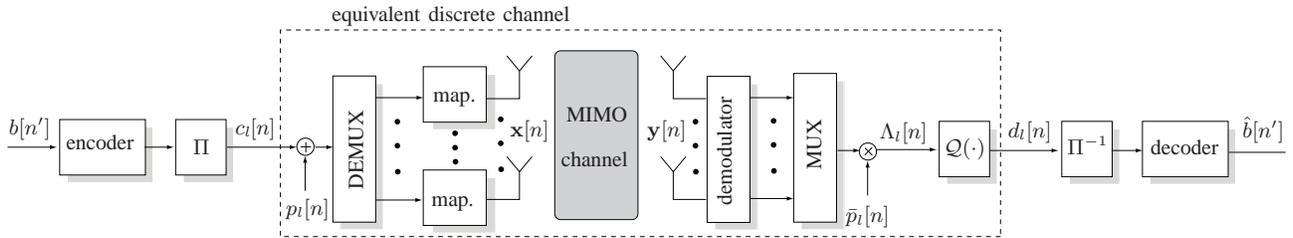}}

\vspace*{-1mm}
\caption{Block diagram of a MIMO-BICM system.}

\label{fig:mimo-bicm}

\vspace*{-4mm}

\end{figure*}

% \subsection{Transmission Model}

We consider a MIMO-BICM system with $\sMt$ transmit antennas and $\sMr$ receive antennas
(SISO-BICM can be viewed as special case with $\sMt\!=\!\sMr\!=\!1$).
% The % architecture of the 
% MIMO-BICM model we consider 
A block diagram is shown in Fig.~\ref{fig:mimo-bicm}.
A sequence of information bits $\sbit[\sq]$ is encoded using an error-correcting code, passed through a {\em bitwise interleaver} 
%\footnote{Note that in contrast to BICM, coded modulation uses a {\em symbol interleaver} and yields a different demodulator metric (cf.\ \cite{caire98}).} 
$\Pi$ and then scrambled by a pseudo-random sequence $p_l[n]$.  The uniformly distributed interleaved and scrambled code bits %$\scbit[\sq]$ 
are demultiplexed into $\sMt$ antenna streams (``layers''). 
In each layer, groups of $\sm$ code bits are mapped to 
% At the $\sk$-th layer the coded bits are subsequently mapped (using Gray labeling \cite{caire98}) onto 
(complex) data symbols $\sxi_\sk[\sn]\!\in\!\sA$, $k=1,\dots,\sMt$; here, $\sA$ denotes the symbol alphabet of size $\sM\!=\!2^\sm$. The transmit vector at symbol time $n$ is given by %\footnote{Superscript $^T$ ($^H$) denotes the (Hermitian) transpose.} %%%%%%%%%%%%%%%%%%%%%%%%%%%%%%%%%%%%%%%%%% 
$\sx[\sn] \define (\sxi_1[\sn]\,\dots\,\sxi_{\sMt}[\sn])^T$
and carries $\slambda\!=\!\sm\sMt$ interleaved code bits $\scj[n]$, $\sj\!=\!1,\dots,\slambda$.
% with zero-mean and unit variance.
 
% For any given time instant $\sn$, the resulting symbol vector $\sx[\sn]\!\define\!(\sxi_1[\sn],\dots,\sxi_{\sMt}[\sn])^T$ is transmitted over a fast Rayleigh fading MIMO channel with $\sMt$ transmit and $\sMr\ge\sMt$ receive antennas to obtain
% Assuming a flat fading MIMO channel, the length-$\sMr$ receive vector 
% $\sr[\sn]$ % \define (\sri_1[\sn]\,\dots\,\sri_{\sMr}[\sn])^T$ 
% is given by 
Assuming flat fading, the length-$\sMr$ receive vector equals
%\cite{mullwein02} (see)
\begin{equation}
\sr[\sn] = \sH[\sn]\hspace*{.3mm} \sx[\sn]+\sw[\sn]. % \qquad \sn=1,\dots,\sN, 
\label{eq:system-model}
\end{equation}
Here, $\sH[\sn]$ is the $\sMr\times\sMt$ MIMO channel matrix
and 
% $\sw[\sn]$ % \define(\swi_1[\sn]\,\dots\,\swi_{\sMr}[\sn])^T$
$\sw[\sn]\sim \CNc({\bf 0}, \sigma^2\Ibf)$
denotes the complex Gaussian noise vector.
% whose elements are i.i.d.\ circularly complex Gaussian 
% , and the components of the noise vector 
% , and the elements of the $\sMr\!\times\!\sMt$ channel matrix $\sH[\sn]$ are modeled as i.i.d.\ circularly symmetric complex Gaussian fading with zero mean and unit variance. The components of the noise vector 
% with zero mean and variance $\ssigw$.  
In the following, %remainder of this paper, 
we will omit the time index $\sn$ to simplify notation. 

%The vector\footnote{Here, $\shA\!\define\!\sA\!\times\!\dots\!\times\!\sA$ denotes the $\sMt$-fold Cartesian product of 
%$\sA$.} %%%%%%%%%%%%%%%%%%%%%%%%%%%%%%%%%%%%%%%%%
%$\sx[\sn] \in \shA$ 
%obeys %the component-wise 
%the power constraint $\sE\{\|\sx[\sn]\|^2\}=\sPt$
% $\sE\{|\sxi_\sk[\sn]|^2\}\!=\!\sPt/\sMt$ such that the total transmit power equals $\sPt$
%($\sE\{\cdot\}$ denotes expectation).
%Assuming a channel matrix with normalized entries, the SNR at each receive antenna equals
% The resulting SNR measured at each receive antenna is then given by %$\sSNR\!=%\!\sE\{\|\sH\sx\|^2\}/\sE\{\|\sw\|^2\}\!\!
%$\sSNR\triangleq\sPt/\ssigw$. In the remainder of this paper, we will omit the time index $\sn$ for convenience. 

% \subsection{Receiver}

At the receiver,
% consists of a demodulator, a de-interleaver,
% , a quantizer 
% and a channel decoder. 
the max-log demodulator calculates % (possibly approximate/quantized) 
LLRs 
% $\sLLRj$ 
for the code bits $c_l$ according to
\cite{mullwein02}
\begin{align}
\sLLRj & 
% \define \log \frac{\sprob(\scj \! = \!1|\sr,\sH)}{\sprob(\scj\!=\!0|\sr,\sH)} \nonumber \\
% & \approx 
= \frac{1}{\ssigw}\!
\bigg[ 
{\underset{\sx\in\shAz}{\min} \|\sr\!-\!\sH\sx\|^2}\,-{\underset{\sx\in\shAo}{\min} \|\sr\!-\!\sH\sx\|^2}
\bigg]. \label{eq:max-log}
\end{align}
%\bee
%\sLLRj \approx \frac{1}{\ssigw}\!\bigg[ {\underset{\sx\in\shAz}{\min} \|\sr-\sH\sx\|^2}\,-{\underset{\sx\in\shAo}{\min} \|\sr-\sH\sx\|^2}\bigg], \label{eq:max-log}
%\eee
%
Here, ${\cal X}_l^b$
% $\shAo$ and $\shAz$ 
denotes the % complementary 
set of transmit vectors for which $c_\sj=b$.
% and $c_\sj=0$, respectively.
% and the max-log approximation \cite{mullwein02} was used.
% \!=\! \frac{\sprob(\scj \! = \!1|\sr,\sH)}{\sprob(\scj\!=\!0|\sr,\sH)}$,
% which are quantized by the quantizer $\Qc(\cdot)$. After deinterleaving, the quantized values are decoded by 
These LLRs (or approximate/quantized versions thereof) are de-scrambled by the sequence $\bar{p}_l[n] = 1\!-\!2p_l[n]$, de-interleaved and used by
the channel decoder to obtain bit estimates $\hat\sbit[\sn]$. % for the bits $\sbit[\sn]$.
%
% \subsubsection{Soft Demodulation}
% Assuming that all transmit vectors are equally likely and 
%
% Using the conditional distribution  $\sr|\sx,\sH \sim \CNc(\sH\sx, \sigma^2\Ibf)$ (cf.\ \eqref{eq:system-model}),
%
The symmetric noise distribution and the use of the scrambler yield %imply 
the symmetries 
$f_{\Lambda}(\xi) = f_{\Lambda}(-\xi)$ and
$f_{\Lambda|c}(\xi|c\!=\!1) = f_{\Lambda|c}(-\xi|c\!=\!0)$
for the (un)conditional LLR distribution.
Hence, knowledge of $f_{\Lambda|c}(\xi|c\!=\!1)$ is sufficient for characterizing $\sLLR$.

\section{LLR Quantization}
\label{sec.quant}

The LLRs in \eqref{eq:max-log} can attain any real value. We next study how to quantize these LLRs. While in practice the demodulator will directly deliver
quantized LLRs, the efficient calculation of quantized LLRs is out of the scope of this paper. 

We consider a $q$-bit quantizer characterized by $K=2^q$ bins
$\Ic_k=[i_{k-1},i_k]$, $k=1,\dots,K$.
We use the convention $i_0=-\infty$, $i_K=\infty$
and assume symmetric bins (this is motivated by the symmetry of the LLR distributions), with
boundaries $i_k$ sorted in ascending order.
The quantizer $\Qc(\cdot)$ maps the LLR $\sLLR_l$ to a discrete LLR $d_l$
according to 
\bee
d_l = \Qc(\sLLR_l) = \lambda_k \quad \text{if} \,\, \sLLR_l \in \Ic_k \, .
\eee
Here, $\lambda_k\in\Ic_k$ is the $k$th quantization level.

% $d \! = \! \Qc(\sLLR) \in \{1,2,\cdots,K\}$.
%To this end, the set of real numbers $\mathbb{R}$ is partitioned into $K = 2^q$ non-overlapping sets, where $q$ denotes the number of bits required for quantization. The first interval is defined by $\Ic_1 = [-\infty;i_1]$, an arbitrary interval defined as $\Ic_k = [i_{k-1};i_k]$, and the last interval is given by $\Ic_K = [i_{K-1};\infty]$. The quantizer outputs the index of the interval the LLR $\sLLR$ is in
%\bee d = \Qc(\sLLR) = k \quad \text{if} \,\, \sLLR \in \Ic_k \, .\eee
%
In the following, we consider the equivalent discrete channel with binary input $c\in\{0,1\}$ and $K$-ary output 
$d\in\{\lambda_1,\dots,\lambda_K\}$. %%% as shown in Fig.~\ref{fig.discrete_channel}. 
Here, $c$ and $d$ are obtained by randomly picking a bit position $l=1,\dots,R_0$ according to a 
uniform distribution. This models a situation where the outer channel code is ``blind'' to the 
bit positions within the symbol labels.
The crossover probabilities 
$p_{bk}=\Pr\{d=\lambda_k | c=b \}=\Pr\{\sLLR\in\Ic_k | c=b \}$
% $\epsilon_{c,k}$ 
of this channel are given by
\bee
p_{bk} = \int_{\Ic_k} f_{\sLLR|c}(\xi|b)\, d\xi,
\eee
where $f_{\sLLR|c}(\xi|b)$ is the conditional probability density function (pdf) of the LLR $\sLLR$ given that $c=b$ (averaged with respect to bit position $l$). 
Note that $\Pr\{d = \lambda_k \} =
\Pr\{\sLLR\in\Ic_k \}= \frac{1}{2}(p_{0k} + p_{1k})$. 
The mutual information (capacity) $I=I(c\,;d)$ of this discrete channel is given by \cite{cover91}
\be\label{eq:cap}
I = \frac{1}{2} \sum_{b=0}^1\sum_{k=1}^K p_{bk} \log_2\! \frac{2 p_{bk}}{p_{0k} + p_{1k}}.
%I = \frac{1}{2} \bigg[ \sum_{k} \epsilon_{-1,k} \log_2 \frac{2 \epsilon_{-1,k}}{\epsilon_{-1,k} + \epsilon_{1,k}} + \sum_{k} \epsilon_{1,k} \log_2 \frac{2 \epsilon_{1,k}}{\epsilon_{-1,k} + \epsilon_{1,k}} \bigg].
\ee
If the LLR distribution $f_{\sLLR|c}(\xi|b)$ and hence the transition probabilities $p_{bk}$
are averaged with respect to the statistics of the physical channel $\sH$ (reflecting fast fading), 
the quantity $I$ describes the ergodic rate achievable over the equivalent channel
(cf.~\cite{tsevis05}).
Otherwise (quasi-static fading), the transition probabilities $p_{bk}$, and thus the rate $I$, change with 
every realization of the channel matrix $\sH$. Here, the probability 
\be\label{eq:outagerate}
p_{\text{out}}(r)=\Pr\{I\le R\}\,,\quad 0 \le R \le R_0
\ee 
characterizes the rate (denoted $R$) versus outage trade-off \cite{tsevis05}.
% , where $r$ allows for setting the rate.
%
%The transition probabilities $p_{bk}$ and hence the capacity $I$ of the equivalent channel depend on the 
%LLR quantizer (i.e., bins $\Ic_k$ or boundaries $i_k$). 

% In \cite{Rave:2009} the quantizer intervals were chosen such that the mutual information $I(c;d)$ was maximized. However for our setup, this optimization becomes very difficult to implement. 

Designing the quantizer to maximize the mutual information $I(c;d)$ appears analytically infeasible in general 
(for BPSK and no fading the solution is given in \cite{Rave:2009}).
Hence, we propose a different approach: 
since $c - \sLLR - d$ is a Markov chain, the data processing inequality implies $I(c\,;d) \leq I(c\,;\sLLR)$. 
In order for $I(c\,;d)$ to be as close as possible to $I(c\,;\sLLR)$ (for fixed $K$), our proposed quantizer maximizes the mutual information $I(\sLLR;d)$. 
With $H(\cdot)$ denoting entropy, it follows that $I(\sLLR;d) = H(d) - H(d|\sLLR)$ and $H(d|\sLLR)=0$ because $d$ is a deterministic function of $\sLLR$.
%
%
%we have $I(c;d)=I(c;\sLLR) - I(c;\sLLR|d)$, with $I(c;\sLLR|d)$ quantifying the information loss of the quantizer. The optimal quantizer minimizes $I(c;\sLLR|d) = H(c|d)-H(c|\sLLR)$ ($H(\cdot)$ denotes entropy); since $H(c|\sLLR)$ is independent of the quantizer, this amounts to minimizing $H(c|d)$, which can be shown to be achieved by a uniform distribution for $d$.
%Since $c - \sLLR - d$ is a Markov chain, the data processing inequality implies $I(c\,;d) \leq I(c\,;\sLLR)$. In order for I(c\,;d) $ to be as close as possible to $I(c\,;\sLLR)$ (for fixed $K$), the optimal quantizer maximizes the mutual information $I(\sLLR;d)$. 
%With $H(\cdot)$ denoting entropy, it follows that $I(\sLLR;d) = H(d) - H(d|\sLLR)$ and $H(d|\sLLR)=0$ because $d$ is a deterministic function of $\sLLR$.
$H(d)$ is maximized by a uniform distribution of $d$ and therefore, the quantizer boundaries $i_k^\star$, $k=1,\dots,K\!-\!1$, have to ensure that
\be\label{eq:quant_design}
\Pr\{d\!=\!\lambda_k\} 
% = \Pr\{\sLLR\in\Ic_k \}
= \frac{p_{0k} + p_{1k}}{2}
= \frac{1}{K}, \quad k=1,\ldots,K \, .
\ee
Using the unconditional cumulative LLR distribution 
$F_\Lambda(\lambda)=
\Pr\{ \Lambda\le\lambda \} =
\frac 1 2\int_{-\infty}^\lambda  \big[f_{\Lambda|c}(\xi|c\!=\!0)+f_{\Lambda|c}(\xi|c\!=\!1)\big]\,d\xi$, 
the optimal boundaries can be obtained by finding the arguments for which 
$F_\Lambda(\lambda)=k/K$, i.e.,
\be\label{eq:optbound}
i_k^\star = F^{-1}_\Lambda\Big(\frac{k}{K}\Big)\,\quad k=1,\dots,K\!-\!1\,.
\ee
%Knowing the distribution $f_{\sLLR|c}$, the optimal intervals $\Ic_k$ can be found by the following sequential  procedure: First the endpoint $i_1$ of the first quantization interval $\Ic_1$ is chosen such that $P(d=1)=\frac{1}{K}$. Using this value, the end point $i_2$ of the second interval $\Ic_2$ is chosen accordingly and so forth, till all $K$ intervals are determined.

We note that for the capacity in \eqref{eq:cap} only the bins (boundaries) are 
relevant, i.e., the actual quantization levels $\lambda_k$ do not influence the achievable rate.
%However, these values are important for soft channel decoding (e.g., by a belief propagation decoder) as they represent reliabilities.
However, these values are important in order to provide the channel decoder (e.g., a belief propagation decoder) with correct reliability information \cite{schwandter:2009}.
% for soft channel decoding (e.g., by a belief propagation decoder) as they represent reliabilities.
% when the channel code is soft decoded by e.g. a belief propagation decoder, 
In view of the equivalent discrete channel, we hence propose to choose the quantization levels as
% the output values of the quantizer should be LLR values, i.e.\ represent the reliability of the decision that the LLR value lies in the $k$th interval. Therefore we propose to use a modified quantizer 
corresponding LLRs
\be\label{eq.quant_mod_out}
\lambda_k^\star 
= \log \frac{\Pr\{c=1|d=\lambda_k\}}{\Pr\{c=0|d=\lambda_k\}} 
% = \log \frac{\Pr\{\sLLR \in \Ic_k|c=1\}}{\Pr\{\sLLR \in \Ic_k|c=0\}} 
= \log \frac{p_{1k}}{p_{0k}}.
\ee
We finally note that $\lambda_k^\star \in {\cal I}_k$.

\rem{%
\begin{figure}[b]

	\psfrag{a1}[c]{$0$}
	\psfrag{a2}[c]{$1$}

	\psfrag{b1}[c]{$\lambda_1$}
	\psfrag{b3}[c]{$\lambda_k$}
	\psfrag{b4}[c]{$\lambda_K$}

	\psfrag{d1}[c]{$p_{0k}$}
	\psfrag{d2}[c]{$p_{1k}$}

	\psfrag{e}[c]{$c$}
	\psfrag{f}{$d$}

	\psfrag{h1}{equivalent modulation}
	\psfrag{h2}{\!\!\!channel}
	
	\psfrag{j}{$\Qc(\cdot)$}
	\psfrag{k}{$\sLLR$}

	\centering
	\resizebox{6cm}{!}{\includegraphics{discrete_channel_v3.eps}}
	%\resizebox{4cm}{!}{\includegraphics{discrete_channel_v2.eps}}

	\caption{Block and transition diagram for the equivalent discrete channel.}
	\label{fig.discrete_channel}
\vspace*{-4mm}
\end{figure}
}

\section{SISO-BICM Systems with BPSK Modulation}
\label{sec.siso}

We next study in more detail the case of a SISO system ($\sMt=\sMr=1$) with BPSK modulation ($R_0=1\,$bpcu)
in Rayleigh fading\footnote{The results in this section also apply to 
the inphase and quadrature phase of SISO systems with Gray-labeled QPSK
and to the two layers of BPSK-modulated $2\!\times\!2$ MIMO systems.}. 
Here, the system model \eqref{eq:system-model} becomes real-valued and 
simplifies to $y = h x + w$, with 
$h\sim{\cal N}(0,1)$, $w\sim{\cal N}(0,\sigma^2/2)$, and $x = 2c-1\in\{-1,1\}$. 
Then, the LLR $\sLLR$ % can be calculated according to 
in \eqref{eq:max-log} equals 
\be\label{eq:LLR}
\sLLR = \frac{h y}{\sigma^2} = \frac{1}{\sigma^2} h(h x + w).
\ee
% Due to 
% Since the distribution $f_{\Lambda|c}(\lambda|c)$ satisfies 
% the symmetry property $f_{\Lambda|c}(\xi|c\!=\!1) = f_{\Lambda|c}(-\xi|c\!=\!0)$,
% it is sufficient to  calculate $f_{\Lambda|c}(\xi|c\!=\!1)$ 
% in order to determine the unconditional LLR distribution $f_{\Lambda}(\xi)$. 

%
% \subsection{Derivation of conditional density function}
%

\subsection{Ergodic Capacity}

Conditioned on $c\!=\!x\!=\!1$, 
the LLR can be rewritten as % quadratic form 
$\sLLR =\frac{1}{\ssigw}\, \zbf^T \!\Abf\, \zbf \, ,$ where
$\zbf = \big(h \; \frac{\sqrt{2}w}{\sigma}\big)^T\sim{\cal N}({\bf 0},{\bf I}) $ and 
\bee
\Abf = % \frac{1}{\ssigw}
%\begin{pmatrix}
%\displaystyle
%	\frac{x}{\ssigw} & \displaystyle\frac{1}{2\sigma} \\
%\displaystyle	\frac{1}{2\sigma} & \displaystyle 0
%\end{pmatrix}\!,
\begin{pmatrix}
\displaystyle
	1   &  \sigma/2 \\
 \sigma/2  &  0
\end{pmatrix}\!.
\eee
Using the eigenvalue decomposition $\Abf=\Ubf\Sigmabf\Ubf^T$,
with $\Ubf$ orthogonal and $\Sigmabf=\text{diag}\{\sigma_1,\sigma_2\}$,
where $\sigma_{1,2}=\frac{1 \pm \sqrt{\rule[-.7mm]{0mm}{2mm}1+\ssigw}}{2}$,
we further obtain
\bee
\sLLR = \frac{1}{\ssigw} \, \tilde\zbf^T \Sigmabf\, \tilde\zbf
= \frac{1}{\ssigw}\Big[\sigma_1 \tilde{z}_1^2 + \sigma_2 \tilde{z}_2^2\Big]\,.
\eee
Here, $\tilde\zbf=\Ubf^T\zbf \sim {\cal N}({\bf 0},{\bf I})$ due to the
orthogonality of $\Ubf$.
Thus, $\sLLR $ is a linear combination of two independent chi-square random variables with one degree of freedom.
%The eigenvalue decomposition of $\Sigmabf^{1/2} \Abf \Sigmabf^{1/2}$ is given by $\Sigmabf^{1/2} \Abf \Sigmabf^{1/2} = \Pbf \Dbf \Pbf^T$, with the unitary matrix $\Pbf$ and $\Dbf=\text{diag} \{ \lambda_1 , \lambda_2 \}$. From the structure of $\Sigmabf^{1/2} \Abf \Sigmabf$ follows $\lambda_1 =-\epsilon$ and $\lambda_2 = 1+\epsilon$ with $\epsilon > 0$. Defining $\ubf = \Pbf^T \ybf$ yields
% \bee z = \ubf^T \Dbf \ubf = \sum_{j=1}^2 \lambda_j u_j^2 \eee
% with $\ubf=(u_1 \,\, u_2)^T \sim \Nc(\zerobf, \Ibf)$. 
% The expression $u^2_j$ is chi-square distributed, and 
The distribution $f_{\Lambda|c}(\xi|c=1)$ can thus be shown to be given 
by (cf.\ \cite{simon:2002})
\be\label{eq:siso_pdf}
f_{\sLLR|c}(\xi|c\!=\!1) = 
\frac{\sigma}{\pi} 
\exp\!\big(-\xi{\sqrt{1+\sigma^2}} \big)\, K_0( |\xi| ),
% \frac{\sigma_w^2}{2} f_{z|c}(\xi|c) \bigg( \frac{\sigma_w^2 \xi}{2} \bigg).
\ee
% \bee
% f_{z|c}(\xi|c=1) = \frac{1}{C} \exp \bigg( - \frac{\lambda_2 - \lambda_1}{4 \lambda_1 \lambda_2} \xi  \bigg) K_0 \bigg( \frac{\lambda_1 + \lambda_2}{4 \lambda_1 \lambda_2} |\xi| \bigg),
% \eee 
% with $C=2\pi \sqrt{|\lambda_1| \lambda_2}$ and 
where $K_0(\cdot)$ denotes the modified Bessel function of the second kind and order 0.

%% i.i.d.\ fast fading

% \subsection{Numerical Results}

\begin{figure}
\centering
% 16:9 Breitband
\includegraphics[scale=0.4]{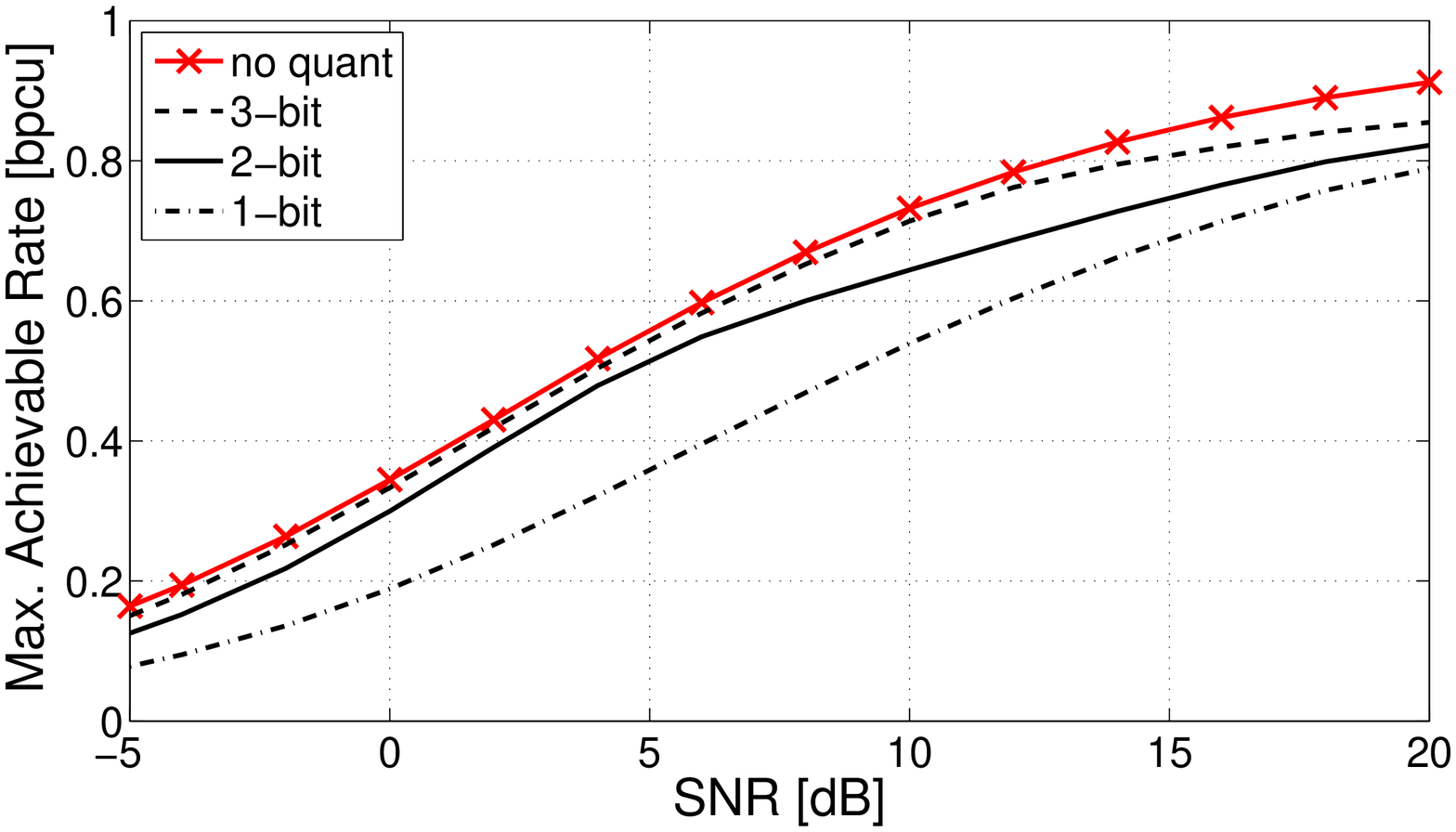}
% 4:3 Ratio
% \includegraphics[scale=0.4]{plots/cap_siso_fading_tv.eps}
\vspace*{-5mm}
\caption{Comparison of ergodic capacity for SISO-BICM with BPSK and different quantizer word-lengths.}
\label{fig:cap_siso}
\vspace*{-2mm}
\end{figure}

% With the use of 
Using \eqref{eq:siso_pdf}, one can determine
the LLR distribution, the LLR quantization (cf.\ \eqref{eq:optbound}),
and the ergodic capacity of the equivalent channel. % given by \eqref{eq:cap}. 
Numerical results for the rate 
in bits per channel use (bpcu) versus SNR 
achievable with our proposed LLR quantizers of different word-length $q$ are shown in Fig.~\ref{fig:cap_siso}.
As a reference, we also show the capacity of non-quantized max-log demodulation (labeled 'no quant').
Hard-output demodulation
(i.e., 1-bit quantization) 
incurs a significant performance loss compared to non-quantized demodulation
(more than 5\,dB at rate $1/2\,$bpcu).
With 2-bit and 3-bit LLR quantization, performance remains within 1\,dB of the non-quantized case
up to rates of approximately $1/2\,$bpcu and $3/4\,$bpcu, respectively. 
% and that the capacity loss of quantization with $2$ and $3$Bits increases with SNR. For a target rate of $0.5$ quantization with $3$Bits is almost sufficient.

\subsection{Outage Capacity}

\begin{figure}
\centering
% 16:9 Breitband
\vspace*{-1mm}
\includegraphics[scale=0.4]{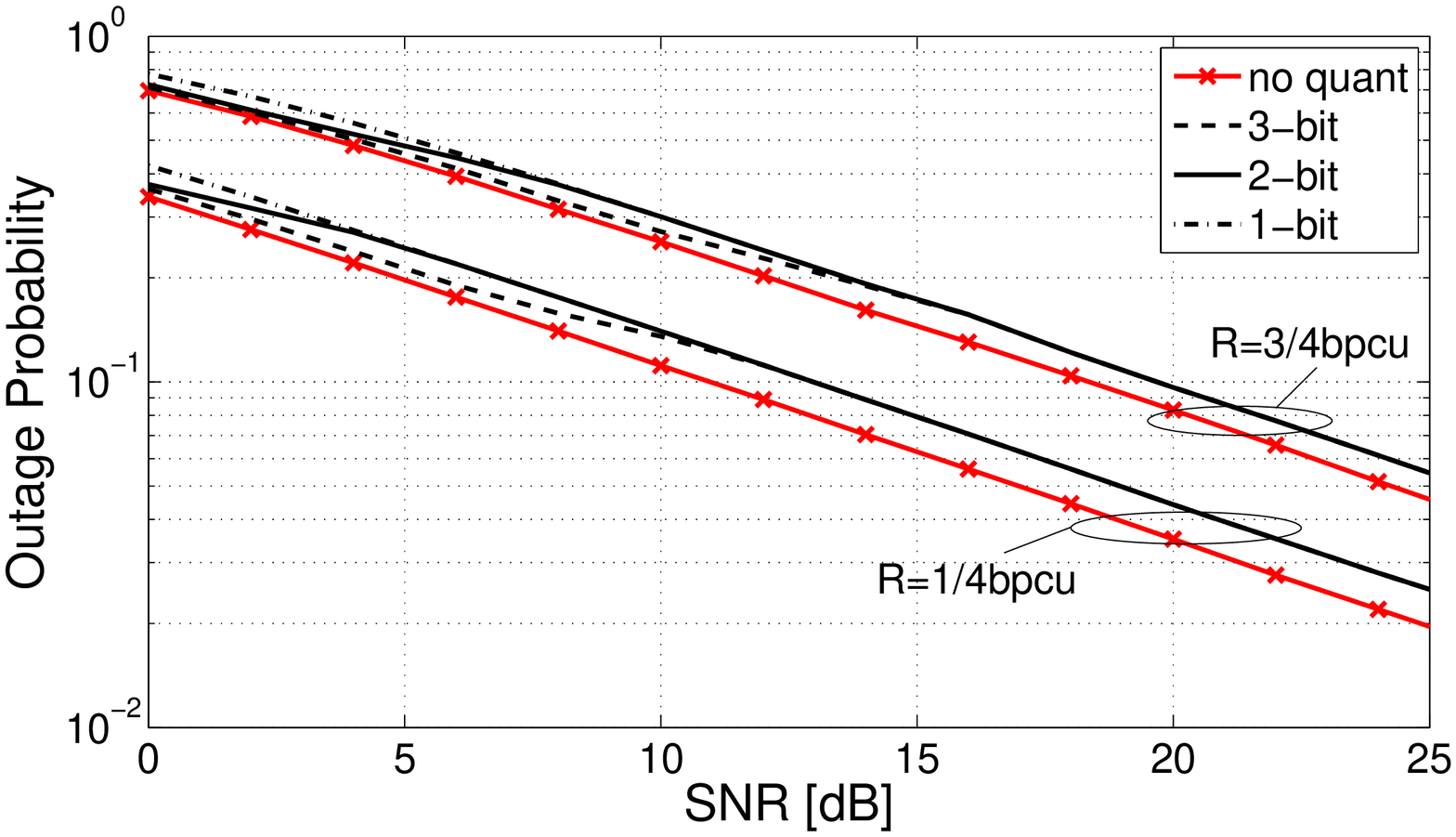}
% 4:3 Ratio
% \includegraphics[scale=0.4]{plots/cap_siso_fading_tv.eps}
\vspace*{-5mm}
\caption{Outage probability for quasi-stationary fading for SISO-BICM with BPSK for rate $\srate\!=\!1/4\,$bpcu and $\srate\!=\!3/4\,$bpcu.}
\label{fig:QS_quant_siso}
\vspace*{-2mm}
\end{figure}

Additionally conditioning on the channel coefficient $h$, it follows straightforwardly
that $\sLLR|c \sim {\cal N}( x\,\gamma, 2\gamma)$ with $\gamma = h^2/\sigma^2$. 
This allows to calculate the transition probabilities of the equivalent channel as
\[
p_{bk} = 
Q\bigg( \frac{i_{k-1} - (2b\!-\!1)\gamma}{\sqrt{2\gamma}} \bigg) - Q\bigg( \frac{i_{k} - (2b\!-\!1)\gamma}{\sqrt{2\gamma}} \bigg) .
\]
The outage probability can thus be evaluated according to \eqref{eq:outagerate}. Numerical results of $p_\text{out}(r)$ versus SNR for quasi-stationary fading with rate $\srate\!=\!1/4\,$bpcu and with $\srate\!=\!3/4\,$bpcu are shown in Fig.~\ref{fig:QS_quant_siso}. 
% It can be seen that 
LLR quantization with more than $2$\,bits is required to offer performance gains at medium-to-high outage probability. At high SNR, the gap between the non-quantized case and
all quantized demodulators ($q=1, 2, 3$) is $2.5$\,dB and $1.5$\,dB for $\srate\!=\!1/4\,$bpcu and $\srate\!=\!3/4\,$bpcu, respectively. Here, $q>3$
% even more bits for quantization are needed 
is required to close this gap and to reach outage probabilities close to the non-quantized case. 
% Furthermore, the asymptotic slopes of these curves show a diversity order of 1 for all quantizer word-lengths.

\section{MIMO Systems and Higher-Order Modulation}
\label{sec.mimo}

In the following, we investigate LLR quantization for MIMO systems and higher-order constellations. 
% To the authors' best knowledge, an 
Since in this case analytical expressions for the LLR distribution are hard to obtain in general,
% is not known for general MIMO channels and higher order modulation, therefore 
the remaining discussion is based exclusively on numerical results.
For the capacity results in this section, 
we used empirical LLR distributions obtained from Monte-Carlo simulations
% \subsection{Quantization Intervals}
to determine the bins ${\cal I}_k$ such that
$\Pr\{\sLLR\in {\cal I}_k\}=1/K$ (cf.~\eqref{eq:quant_design}).
In the remainder of the paper, we will consider a
$2\!\times\!2$ MIMO system with Gray-labeled $16$-QAM modulation (here, $\slambda\!=\!8\,$bpcu). 

% Based on the symmetry assumption $f_{\sLLR|c}(\xi|c=1) = f_{\sLLR|c}(-\xi|c=0)$, the bins were furthermore designed to be symmetric about the origin.

%\begin{itemize}
%\item The demodulator either outputs quantized LLRs directly or a quantizer cares for quantization of the continuous LLRs delivered at the demodulator output.
%\item The continuous LLRs $\sLLRj$ are quantized using $\sQlev$ quantization bins defined by the edge values $\sedge\!=\!(-\infty,\squant_{-(\sQlev\!-\!2)/2},\dots,\squant_{-1},0,\squant_{1},\dots,\squant_{(\sQlev\!-\!2)/2},\infty)^T$. The edge values are chosen  bins. Here, we choose the edge values $\squant_\si$  (needed in the description of Fig.~\ref{fig:SNR_quant}).
%\item In general, the channel decoder is particularly sensitive to actual LLR values with correct reliability information (cite Stefan-ICASSP paper?). Incorrect reliability information passed to the channel decoder may prevent the system from approaching capacity.
%The corresponding quantization codebook can be obtained as
%\begin{equation} \label{eq:codebook}
%\sLLRqi = \frac{\sprob(\sedge_\si<\sLLR\le\sedge_{\si+1}|\scbit=1)}{\sprob(\sedge_\si<\sLLR\le\sedge_{\si+1}|\scbit=0)}, \quad \forall \si=1,\dots,\sQlev\,,   
%\end{equation}
%where $\sedgei\!=\!(\sedge)_\si$ denotes the $\si$th element of the edge vector $\sedge$. 
%\item remap-value codebook is also symmetric
%\item describe look-up table generation
%\end{itemize}

\subsection{Ergodic Capacity}
\begin{figure}
\centering
% 16:9 Breitband
\includegraphics[scale=0.4]{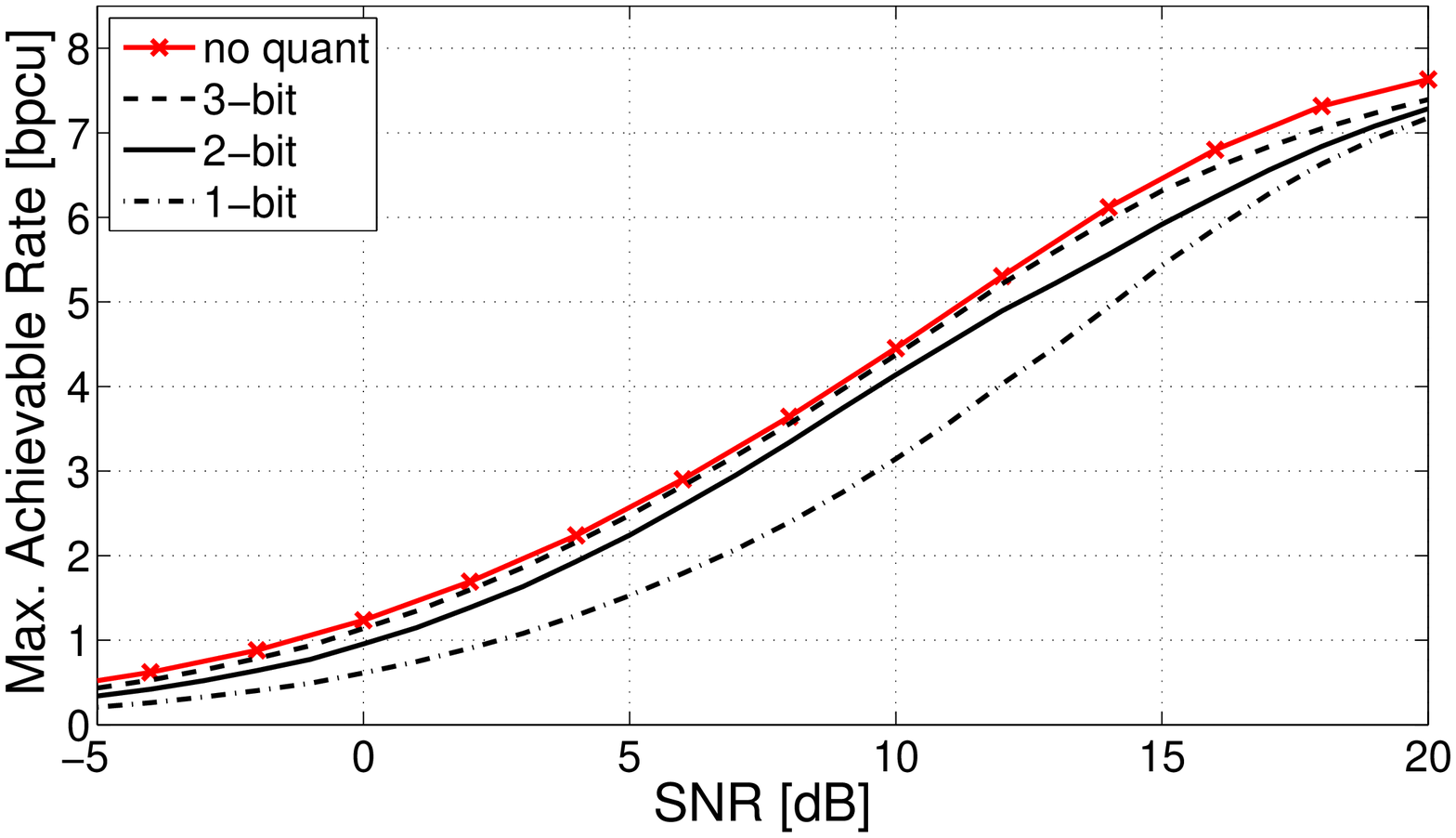}
% 4:3 Ratio
%\includegraphics[scale=0.4]{plots/plot_cap_quant_tv.eps}
\vspace*{-2mm}
\caption{Ergodic capacity of a $2\!\times\!2$ MIMO-BICM system with 
Gray-labeled $16$-QAM for different LLR quantization word-lengths.}
\label{fig:cap_quant}
\vspace*{0mm}
\end{figure}

We evaluated the capacity in \eqref{eq:cap} under the assumption of ergodic 
spatio-temporally i.i.d.\ fast Rayleigh fading for various quantizer word-lengths $q$.
% numbers of bins $K\!=\!2^{\sqbits}$. 
%; here, $\sqbits$ denotes the number of bits required to transmit the quantized signal.. 
%
To this end, we estimated the transition probabilities $p_{bk}$ by means of Monte-Carlo simulations
after having determined the optimal bins based on $10^5$ channel realizations. 
Fig.~\ref{fig:cap_quant} shows the results obtained.
% quantized max-log demodulation in bits per channel use (bpcu) versus SNR for 
% The bins were evaluated over $10^5$ fading realizations. 
It can be seen % in Fig.\,\ref{fig:cap_quant} 
that the extreme case of $1$-bit quantization
% \footnote{We note that the output obtained by max-log demodulation with $1$-bit quantization equals the bit-output resulting from hard-symbol maximum likelihood (ML) detection (cf.~\cite{Fertl:2008aa}).} 
yields a considerable performance loss in comparison to the non-quantized case
(e.g., $3$\,dB SNR loss at $4$\,bpcu). Increasing the number of quantization levels reduces this gap significantly
(to $0.5$\,dB and $0.1$\,dB for $2$-bit and $3$-bit quantization, respectively, at $4$\,bpcu). 
Even though at higher rates slightly increasing SNR gaps are observed,
these results suggest that $3$-bit LLR quantization is sufficient for practical purposes.

\begin{figure}
\centering
% 16:9 Breitband
%\psfrag{D}{$i_3$}
\includegraphics[scale=0.39]{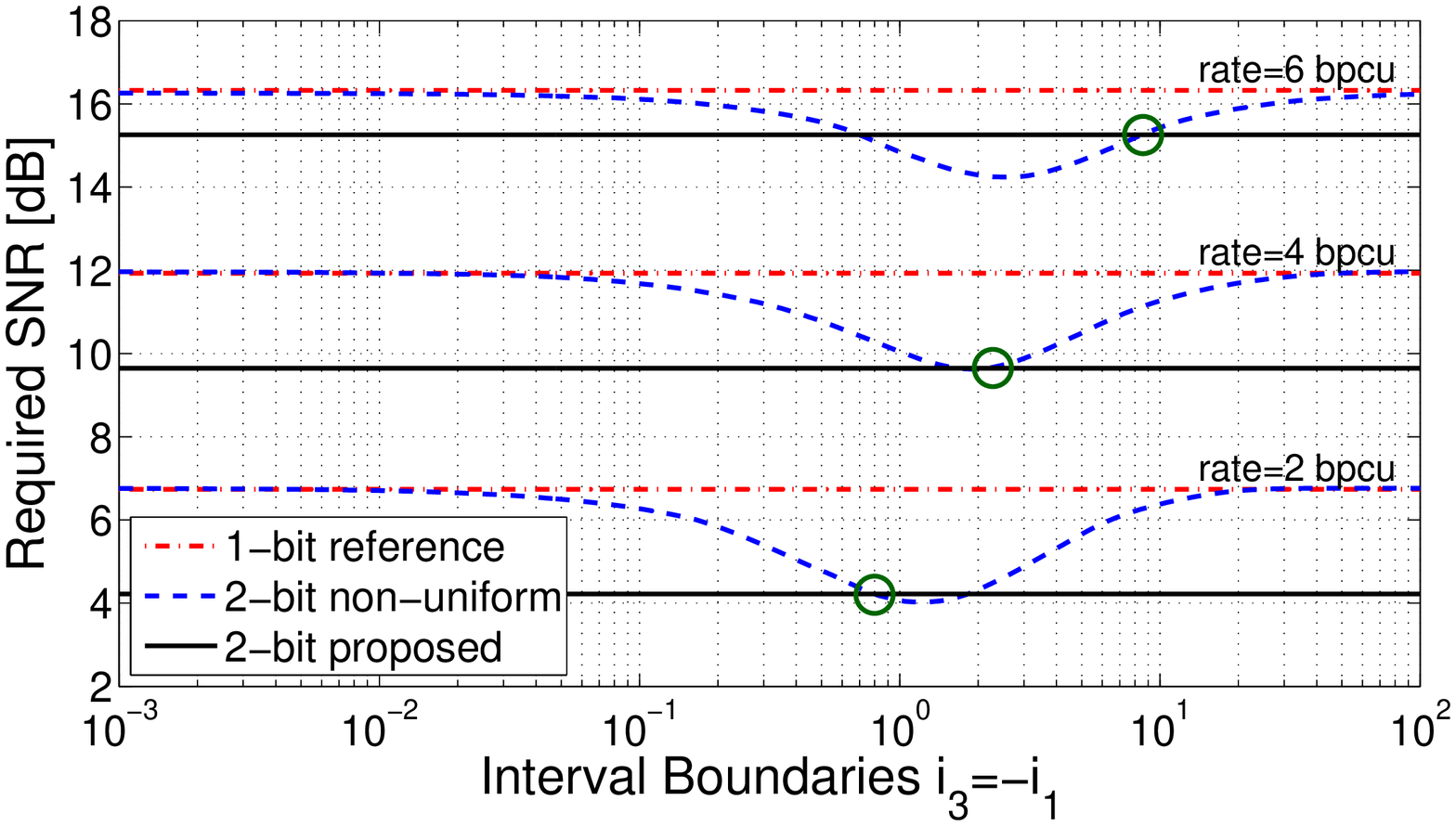}
% 4:3 Ratio
%\includegraphics[scale=0.4]{plots/plot_SNR_vs_quant_tv.eps}
\vspace*{-2mm}
\caption{SNR required for a target rate of $2,4,6$\,bpcu vs.\, quantizer boundary $i_3$ of a 2-bit LLR quantizer 
($2\!\times\!2$ MIMO-BICM system with Gray-labeled $16$-QAM). Boundaries $i_3$ of proposed quantizer  are marked by green dots.}
\label{fig:SNR_quant}
\vspace*{-2mm}
\end{figure}

In order to illustrate the impact of the LLR quantizer design on ergodic capacity, 
we consider the same $2\!\times\!2$ MIMO-BICM system with Gray-labeled $16$-QAM modulation for
various 2-bit (i.e., 4-level) LLR quantizers. Since symmetric quantization here amounts to
$i_2=0$ and $i_1=-i_3$, the boundary $i_3$ is sufficient to index all quantizers in this case.
Fig.~\ref{fig:SNR_quant} plots the SNR required to achieve target rates of 2\,bpcu, 4\,bpcu and 8\,bpcu versus $i_3$, respectively.
As a reference we also show the required SNR using a 2-bit quantizer with uniform distribution and the required SNR for $1$-bit quantization (hard demodulation).
It can be seen that for rates of 2\,bpcu and 4\,bpcu the 2-bit quantizer with uniform distribution requires the same SNR as the 2-bit quantizer with optimal choice of $i_3$ (this is the quantizer proposed in \cite{Rave:2009}). For rates of 6\,bpcu the SNR loss of the 2-bit quantizer with uniform distribution is about 1\,dB compared to the optimal quantizer.

%The optimum quantizer boundary in this case is $i_3^\star \approx 2$. When the boundary 
%$i_3$ deviates from this optimal value, the required SNR increases rapidly and,
%for very small or very large $i_3$ approaches the case of 1-bit quantization,
%requiring an SNR that is about 2.4\,dB higher than with the optimal quantizer.

%In order to illustrate the impact of the LLR quantizer design on ergodic capacity, 
%we consider the same $2\!\times\!2$ MIMO-BICM system with Gray-labeled $16$-QAM modulation for
%various 2-bit (i.e., 4-level) LLR quantizers. Since symmetric quantization here amounts to
%$i_2=0$ and $i_1=-i_3$, the boundary $i_3$ is sufficient to index all quantizers in this case.
%Fig.~\ref{fig:SNR_quant} plots the minimum SNR required to achieve 4\,bpcu versus $i_3$.
%As a reference we also show the required SNR for $1$-bit quantization (hard demodulation).
%The optimum quantizer boundary in this case is $i_3^\star \approx 2$. When the boundary 
%$i_3$ deviates from this optimal value, the required SNR increases rapidly and,
%for very small or very large $i_3$ approaches the case of 1-bit quantization,
%requiring an SNR that is about 2.4\,dB higher than with the optimal quantizer.

% It can be seen that the choice of the edge value is crucial and that deviations from the ``optimal'' choice of the interval limit %(i.e., obtained by uniform distribution of the LLRs) 
% can result in a SNR loss of up to $2.4$\,dB. In fact, for large deviations from the optimal value the performance essentially ends up in $1$\,bit quantization (cf.~Fig.~\ref{fig:cap_quant}). This is quite intuitive since for the case of a very large (or equivalently a very small) edge value $i_1$, we essentially approach the optimal intervals for $1$\,bit quantization.

%\begin{figure*}
%\centering
% 16:9 Breitband
%\includegraphics[scale=0.4]{plots/plot_quant_val.eps}
% 4:3 Ratio
%\subfigure[]{\includegraphics[scale=0.38]{plots/plot_SNR_vs_quant_sq.eps}}
%\subfigure[]{\includegraphics[scale=0.38]{plots/plot_ber_quant_tv.eps}}
%\subfigure[]{\includegraphics[scale=0.38]{plots/plot_bergap_sq.eps}}
%\caption{Robustness of quantization values using $2$-bit quantization for $2\!\times\!2$ MIMO with $16$-QAM using Gray labeling.}
%\label{fig:ber}
%\end{figure*}

\subsection{Outage Capacity}
\begin{figure}
%\centering
% 16:9 Breitband
\includegraphics[scale=0.39]{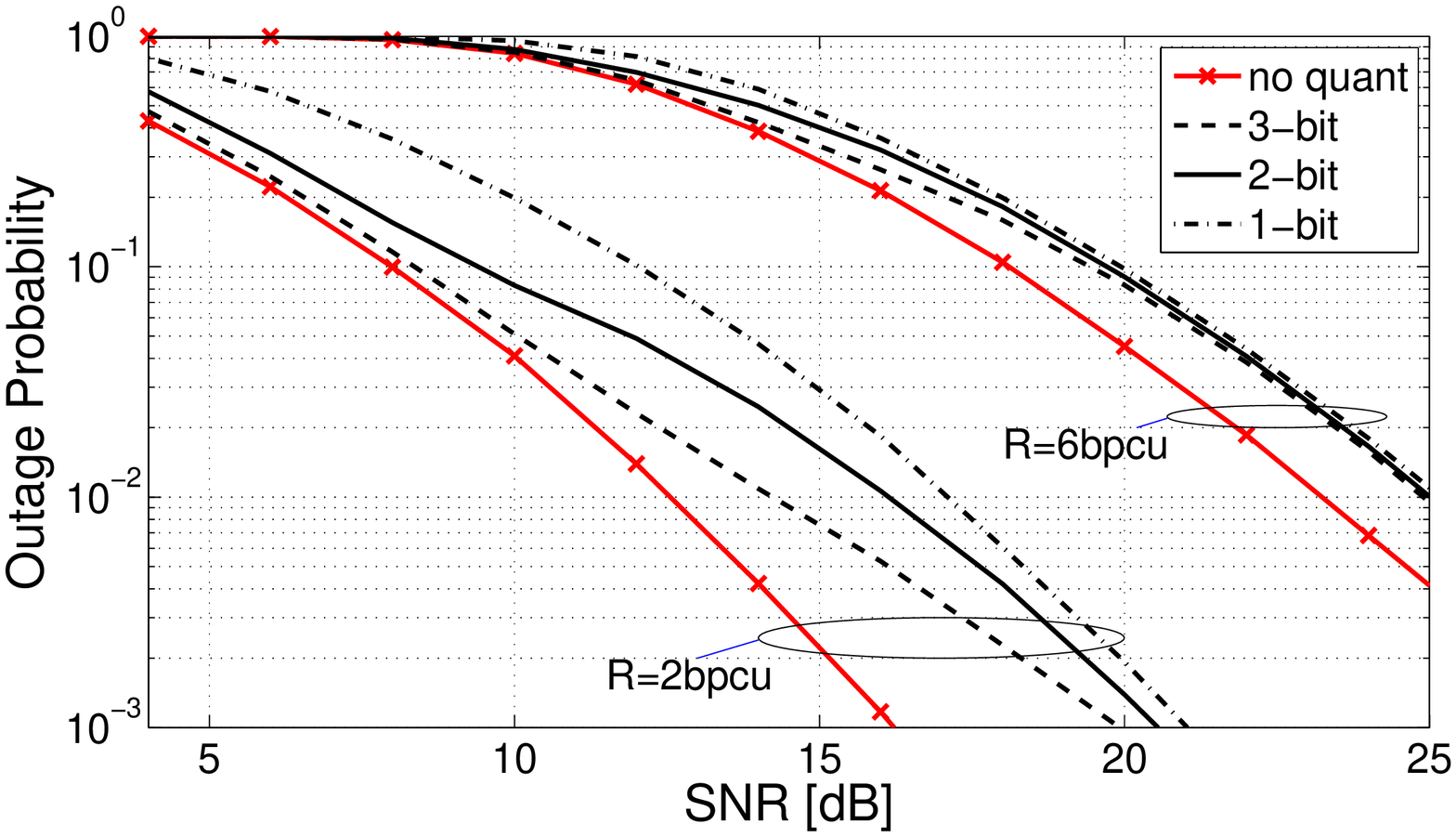}
% 4:3 Ratio
%\includegraphics[scale=0.4]{plots/plot_QS_quant_tv.eps}
\vspace*{-6mm}
\caption{Outage probability for quasi-stationary fading for $2\!\times\!2$ MIMO with $16$-QAM using Gray labeling for rate $\srate\!=\!2\,$bpcu and $\srate\!=\!6\,$bpcu.}
\label{fig:QS_quant}
\vspace*{-2mm}
\end{figure}

% {\em Outage Probability.} 
We next provide numerical results for the outage probability in \eqref{eq:outagerate} 
for quasi-static fading. 
% Here, we use the  as performance measure.
% In this case, the notion of capacity does not represent a relevant performance measure since one no longer codes over different channel realizations \cite{tsevis05}. Therefore we analyze the outage probability defined as
% \begin{equation} \sPout=\sProb(\sR<\srate\,\slambda) \end{equation}
% which essentially equals the probability that the mutual information measure $\sR$ in \eqref{eq:cap} falls below a certain target rate. %; here, $0\!\le\!\srate\!\le\!1$ allows for setting the target rate. 
% Note that in correspondence with the quasi-static fading assumption, $\sR$ in \eqref{eq:cap} is {\em no longer} averaged over different channel realizations, i.e., the transition probabilities $\epsilon_{c,k}$ required in \eqref{eq:cap} are estimated over a transmission block that belongs to the same fading realization. 
% We again consider a $2\!\times\!2$ MIMO system with $16$-QAM employing Gray labeling. 
% \vspace*{1mm} {\em Simulation Results.}
Fig.\,\ref{fig:QS_quant} shows the outage probability $p_\text{out}(r)$ versus SNR %$\sSNR$ 
for different quantizer word-lengths and $R\!=\!2\,$bpcu and $R\!=\!6\,$bpcu. %(i.e., 2 and 6\,bpcu, respectively). 
% The results are presented for a target rate of $\srate\!=\!1/4$ and $\srate\!=\!3/4$. 
%The outage probability $\sPout$ was obtained over $10^5$ fading realizations and for each channel realization $\sR$ %in \eqref{eq:measure-exact} was measured by transmitting a block of $10^4$ symbol vectors. 
%In this case the necessary interval limits of the quantizer, were taken from a precomputed look-up table and have been computed for an i.i.d.\,fast Rayleigh fading channel using $10^5$ fading realizations.
From the asymptotic slopes of these curves it is seen that the diversity order equals 2 in all cases. 
% It is seen that the diversity gain is independent of the number of quantization bins and the target rate. 
For $\srate\!=\!2\,$bpcu and $\srate\!=\!6\,$bpcu, 
hard demodulation ($q=1$)
% the $1\,$-bit quantizer 
is respectively $4.8\,$dB and $1.8\,$dB away from the non-quantized case at high SNR.
LLR quantization with 2 and 3 bits performs only slightly better at very low outage probability,
but offer significant gains at medium-to-high outage probability. 
For $R=2\,$bpcu, the SNR loss of LLR quantization with 1, 2, and 3 bits at $p_\text{out}\!=\!10^{-1}$
equals 4\,dB, 1.4\,dB, and 0.4\,dB, respectively.

% while for higher outage probability, the curves with $2$ and $3$Bit quantization are quite close to the unquantized curve. For lower outage probability these curves bend away from the unquantized curve. For higher rates ($\srate\!=\!3/4$), the behaviour is quite similar, only the gaps are smaller.

%At low rates ($\srate\!=\!1/4$) the extreme case of $1$-bit quantization shows a significant SNR loss in comparison to $3$-bit quantization, e.g., about $4$\,dB for $\sPout\!=\!10^{-2}$. However, for high SNRs these gaps tend to decrease. We note that in this plot $3$-bit quantization serves as a reference, since it already allows to closely approach the performance obtained without quantization.
%At high rates  the performance gaps between different number of bins vanish and the curves essentially coincide for outage probabilities. %, i.e., roughly within $1.5$\,dB. 
%We thus conclude that for high rates a small number of bins can be used for quantization than for low rates. ({\em In fact, this means that hard ML detection should be used instead of max-log demodulation (without quantization)!!! However it is not clear if this may also be the case for other antenna configurations!!!!!})

%\begin{figure*}

%\subfigure[]{ \includegraphics[scale=0.26]{plots/plot_QS_quant.eps} }
%\subfigure[]{ \includegraphics[scale=0.26]{plots/plot_ber_quant.eps} }
%\subfigure[]{ \includegraphics[scale=0.26]{plots/plot_bergap.eps} }

%\end{figure*}

\section{Estimation of Quantization Parameters} \label{sec:estimator}

The computation of the quantization boundaries $i_k^\star$
and quantization levels $\lambda_k^\star$ according to 
\eqref{eq:optbound} and
\eqref{eq.quant_mod_out}, respectively, requires
% the transition probabilities of the equivalent discrete channel and thus in turn 
the LLR distributions 
% From \eqref{eq.quant_mod} and it follows that the calculation of the remap values requires knowledge of the conditional density 
$f_{\sLLR}(\xi)$ and
$f_{\sLLR|c}(\xi|c)$, which in general are unknown.
We thus address on-the-fly estimation of the quantization parameters.
%  which to the author's best knowledge is unknown for MIMO demappers with higher-order modulation. Therefore 
The boundaries can be estimated by using an empirical estimate of the unconditional LLR distribution $F_{\sLLR}(\xi)$, which can be obtained from a reasonable number of 
non-quantized LLRs. % without knowing the associated code bit. 

In contrast, determining the quantization levels $\lambda_k^\star$ by estimating $f_{\sLLR|c}(\xi|c)$ 
is more difficult since the code bits are unknown at the receiver. Hence, we propose to % a method for estimating the $\lambda_k^\star$ by 
use the following simple parametric model, which is motivated by numerical results
for the $2\times 2$ case with 16-QAM (other system parameters may require a
different model):
% At the receiver the model parameters are estimated from the observed values $\sLLR$. 
% Motivated by numerical results, we use the simple model
% results, the conditional density $f(\sLLR|c)$ is modeled by
\vspace*{-.7mm}
\be
\label{eq:pdf_model}
f_{\sLLR|c}(\xi|c\!=\!1) = 
\begin{cases}
\frac{\alpha\beta}{\alpha+\beta} \exp (\alpha \xi) &  \xi   <  0, \\
\frac{\alpha\beta}{\alpha+\beta} \exp (-\beta  \xi) &  \xi \geq 0.
\end{cases}
\ee
% (we assume $f_{\sLLR|c}(\xi|c\!=\!0)=f_{\sLLR|c}(-\xi|c\!=\!1)$). 
% with $c = \frac{1}{1/\lambda_l + 1/\lambda_r}$. %Note that $f_{\sLLR}(\xi) = 1/2 (f_{\sLLR|c}(\xi|c=0) + f_{\sLLR|c}(\xi|c=1))$.
%

\vspace*{-.7mm}\noindent
To estimate the two parameters $\alpha>0$ and $\beta>0$, we choose two bins 
$\bar \Ic_{1}$ %=[A_{1,2} \,\, B_{1,2}]$ 
and $\bar \Ic_{2}$ 
and use the non-quantized LLRs $\sLLR$ to obtain empirical estimates $\hat{P}_i$, $i=1,2$, of the probabilities 
% $\hat P_{i} = \Pr\{ \sLLR \in \bar \Ic_{i}\}$. From the model \eqref{eq:pdf_model} these probabilities can be obtained as 
\bee
P_i(\alpha,\beta)=\Pr\{\sLLR \in \bar\Ic_{i}\} = \int_{\bar \Ic_{i}} f_{\sLLR}(\xi) \,d\xi \, ,
\eee

\vspace*{-1mm}\noindent
with $f_{\sLLR}(\xi) = \big[f_{\sLLR|c}(\xi|c\!=\!0) + f_{\sLLR|c}(\xi|c\!=\!1)\big]/2$.
The system of equations $P_i(\alpha,\beta)=\hat{P}_i$ can then be solved numerically to obtain
estimates of $\alpha$ and $\beta$. The transition probabilities of the equivalent channel
and the quantization levels are then computed based on \eqref{eq:pdf_model} using the estimates
of $\alpha$ and $\beta$.
% Note that $P(\sLLR \in \Ic_{1,2})$ depend on $\lambda_r$ and $\lambda_l$ and by comparing $\hat P_{1,2}$ with $P(\sLLR \in \Ic_{1,2})$ the two parameters can be obtained numerically.
% Note that for other system configurations a different parametric model for $f_{\sLLR|c}(\xi|c\!=\!1)$ may be required. 

\section{Numerical BER Results} \label{ssec:ber}

To verify the capacity results, we performed BER simulations for SISO- and MIMO-BICM systems
in ergodic Rayleigh fast fading. 
The channel code was a regular LDPC
% {\em low-density parity check (LDPC)} 
code\footnote{The LDPC code was designed 
using the EPFL web-tool at 
{\texttt{http://lthcwww.epfl.ch/research/ldpcopt}}.}
with rate 1/2 and block length 64000.

\subsection{SISO-BICM}
\begin{figure}
\centering
% 16:9 Breitband
\hspace*{-4mm}
\includegraphics[scale=0.39]{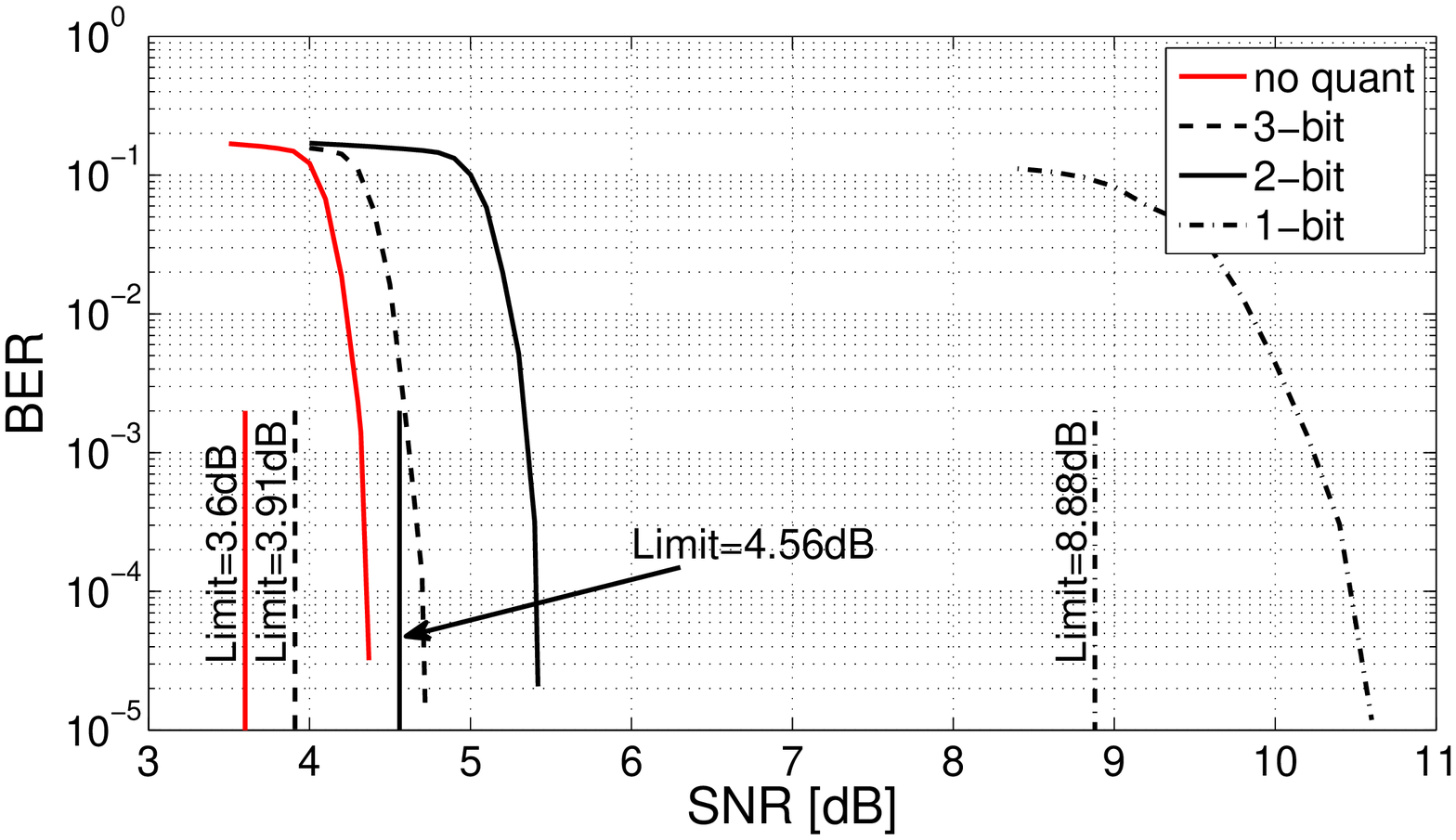}
% 4:3 Ratio
%\includegraphics[scale=0.4]{plots/ber_siso_fading_tv.eps}
\vspace*{-2mm}
\caption{BER performance for a rate-1/2 LDPC coded 
SISO-BICM system with BPSK and different LLR quantization word-lengths.}
\label{fig:ber_siso}
\vspace*{-1mm}
\end{figure}

We first consider a BPSK-modulated SISO-BICM system 
with LLR quantizers
designed using the analytical results from Section \ref{sec.siso}.
Fig.~\ref{fig:ber_siso} shows the BER for 
our proposed LLR quantizers
with 
different word-length together with the theoretical SNR thresholds (obtained from Fig.~\ref{fig:cap_siso}). All BER curves are reasonably close to the 
% corresponding 
respective SNR thresholds
(obtained from Fig.~\ref{fig:cap_siso} and indicated by vertical lines). 
The gaps of 1-bit, 2-bit, and 3-bit LLR quantization to the non-quantized case respectively equal
6.2\,dB, 1.1\,dB, and 0.4\,dB.
% within $1$dB of the capacity limits, only the curve corresponding to a quantization with $1$bit shows a larger gap. This is due to the fact, that the used LDPC code was optimized for AWGN channels which performs worse in this quasi BPSK setting.

\subsection{MIMO-BICM}
\begin{figure}
\centering
% 16:9 Breitband
\includegraphics[scale=0.39]{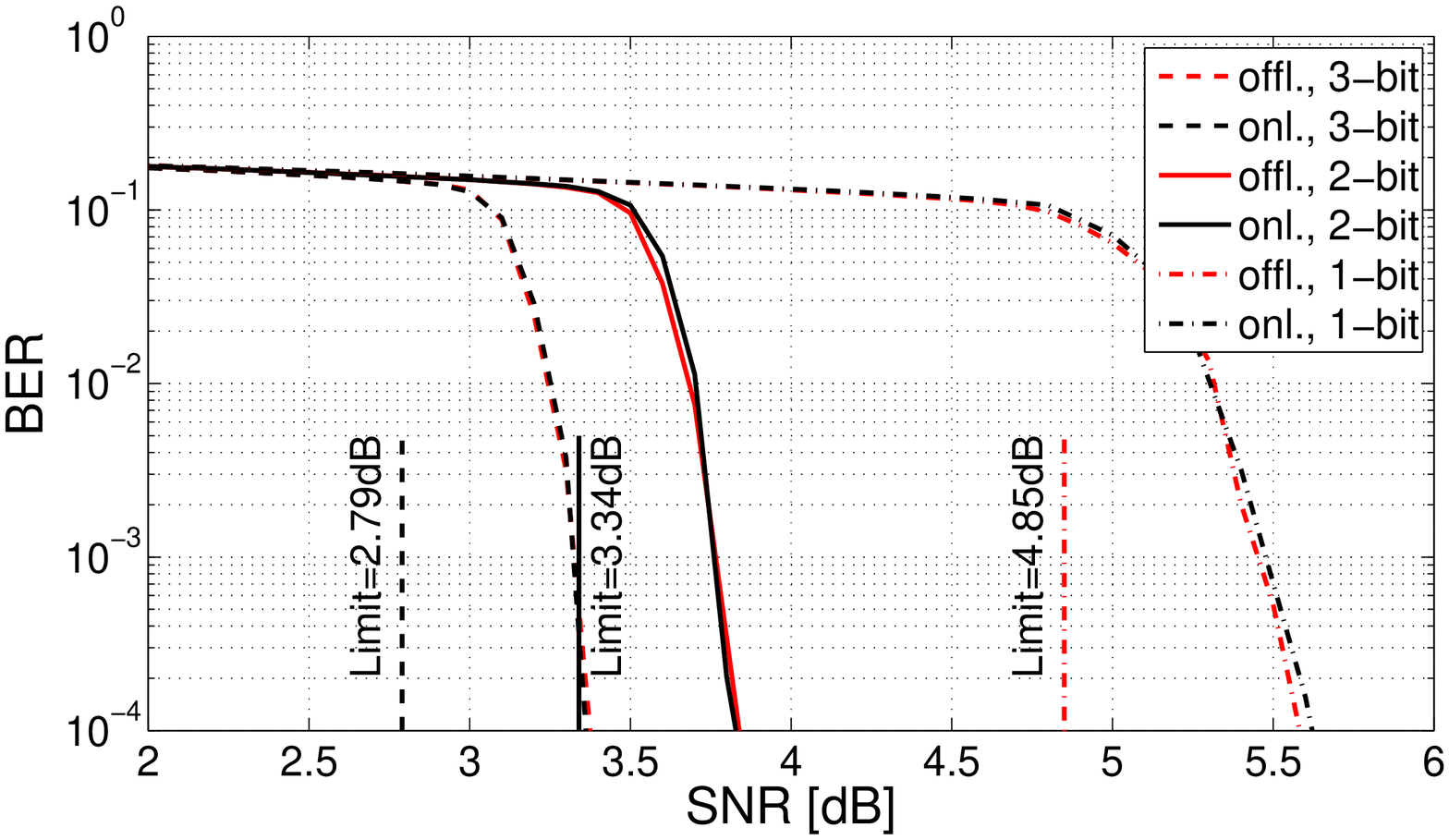}
% 4:3 Ratio
%\includegraphics[scale=0.4]{plots/plot_ber_quant_tv.eps}
\vspace*{-2mm}
\caption{BER performance for a rate-1/2 LDPC coded 
$2\!\times\!2$ MIMO system with Gray-labeled 16-QAM
and different LLR quantization word-lengths.}
\label{fig:ber_quant}
\vspace*{-3mm}
\end{figure}

% {\em BER Performance.}
Fig.\ \ref{fig:ber_quant} shows two (strongly overlapping) sets of BER curves % versus SNR 
for the $2\!\times\!2$ MIMO-BICM system with Gray-labeled 16-QAM and
different LLR quantization word-lengths.
One set of curves (labeled `offl.')
pertains to an offline design of the LLR quantizer% 
%based on Monte-Carlo simulations using $10^5$ fading realizations
,
whereas the other set (labeled `onl.')
estimates the quantization parameters on-the-fly % for each block 
according to Section \ref{sec:estimator}. 
% The bins $\bar \Ic_{1}$ %=[A_{1,2} \,\, B_{1,2}]$  and $\bar \Ic_{2}$ were chosen heuristically. 

% Specifically, we plot the performance results for the case of taking the edge- and remap-values from a precomputed look-up table (labeled 'LT'). %Here, the corresponding values in the look-up table were computed over $10^5$ fading realizations for the channel parameters of interest. 
% Additionally, we show results for the case of computing the quantization bins on-the-fly for each codeword (of length $64000$) and estimating the remapped LLR values with the estimator proposed in \ref{sec:estimator} (labeled 'est'). The capacity limits in terms of minimum SNR required to support rate $1/2$ (cf.~Fig.~\ref{fig:cap_quant}) are also indicated.

The gap to the theoretical SNR thresholds 
(obtained from Fig.~\ref{fig:cap_quant} and indicated by vertical lines)
equals $0.6$\,dB for 
$3$-bit and $2$-bit quantization 
and $1$\,dB for $1$-bit quantization (hard demodulation). 
Furthermore, the proposed on-the-fly estimator for the LLR quantizer parameters
% remap-values 
performs extremely well in this setup (virtually indistinguishable from the offline design).
% coincides with the results obtained with the look-up table. 
% Since the estimator was specifically designed for the discussed system setup, we note that 

%
% \vspace*{1mm}

\begin{figure}
\centering
% 16:9 Breitband
\vspace*{1.5mm}
\includegraphics[scale=0.39]{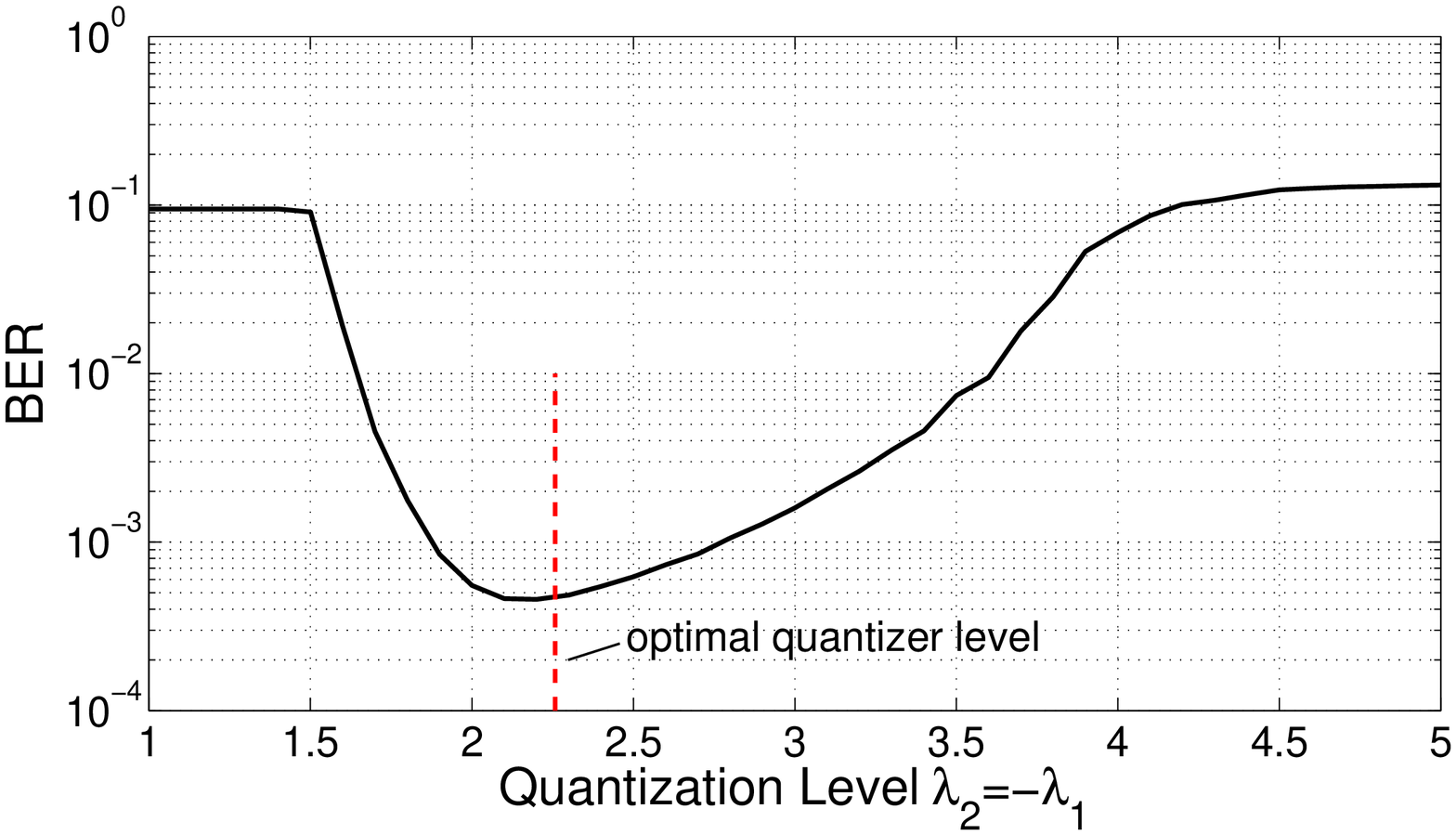}
% 4:3 Ratio
%\includegraphics[scale=0.4]{plots/plot_bergap_tv.eps}
\vspace{-3.7mm}
\caption{BER versus quantization level for $1$-bit quantization at an SNR of $12.8$\,dB using a rate-$1/2$ LDPC code ($2\!\times\!2$ MIMO, $16$-QAM, Gray labeling).}
\label{fig:bergap}
\vspace*{-4mm}
\end{figure}

To illustrate the importance of the correct choice of the LLR quantization levels,
Fig.~\ref{fig:bergap} shows BER versus quantization level $\lambda_2=-\lambda_1$ 
for the same MIMO system as before with 1-bit LLR quantization at an SNR of $12.8\,$dB.
% {\em Choice of Remap-Values.}
% Next we investigate how the performance of the channel decoder is affected by the actual choice of the remap-value, i.e., if the decoder is fed with incorrect reliability information. To this end we show the BER versus remap-value for $1$-bit quantization for a SNR of $12.8$\,dB in Fig.~\ref{fig:bergap}. %Here, the SNR was chosen such that an ``optimal'' choice of the remap-value according to \eqref{eq.quant_mod} results in a low BER of less than $10^{-3}$ (cf.~Fig.~\ref{fig:ber_quant}). 
Here, the optimal quantizer level $\lambda_2^\star=2.26$
(indicated % in Fig.~\ref{fig:bergap} 
by a dashed vertical line)
achieves a BER of $4.5\cdot 10^{-4}$.
% remap-value attained a value of $2.26$ %and was obtained by evaluating \eqref{eq.quant_mod} over $10^5$ fading realizations (indicated in Fig.~\ref{fig:bergap} by a dashed line). 
%
It is seen that the BER achieved by the belief propagation decoder is quite sensitive to 
the choice of $\lambda_2$; 
for $\lambda_2\le 1.5$ or $\lambda_2\ge 4.3$,
BER has deteriorated 
to about $10^{-1}$
(i.e., by more than 2 orders of magnitude).
% incorrect reliability information. If the remap-value is chosen smaller than the optimal choice, the BER performance degrades significantly. For larger remap-values a similar behavior can be observed, although not as pronounced as for smaller values. 

\vspace*{-1mm}
\section{Conclusion}\label{sec.conclusion}
\vspace*{-1mm}

We considered bit-interleaved coded modulation systems with demodulators providing quantized 
log-likelihood ratios (LLR). We provided design rules which lead to easily implementable LLR quantizers and studied the information rates of the equivalent discrete channel in the ergodic and outage regime. Numerical results for capacity and bit error rate showed that LLR quantization using a small number of bits is often sufficient.
% but optimal design of the quantization parameters is crucial. To that end, 
We also proposed simple procedures to estimate the quantizer parameters during 
data transmission.

\vspace*{-1mm}
\section*{Acknowledgements}
\vspace*{-1mm}
The authors are grateful to Jossy Sayir for drawing their attention to the 
problem of quantized soft information and to Joakim Jald\'en for helpful comments.
This work was supported by
the STREP project MASCOT (IST-026905), 
% and 
the Network of Excellence NEWCOM++ (IST-216715),
% within the Sixth Framework Programme of the European Commision,
and by the FWF Grant S10606 ``Information Networks''.
%  within the National Research Network SISE.

\bibliographystyle{ieeetr}
\bibliography{tf-zentral,FPbib}

\end{document}